# An Ethereum-compatible blockchain that explicates and ensures design-level safety properties for smart contracts

(Working draft)

Nikolaj Bjørner, Shuo Chen, Yang Chen, Zhongxin Guo
Microsoft Research

Peng Liu, Nanqing Luo
The Pennsylvania State University

**Abstract**

*Smart contracts are crucial elements of decentralized technologies, but they face significant obstacles to trustworthiness due to security bugs and trapdoors. To address the core issue, we propose a technology that enables programmers to focus on design-level properties rather than specific low-level attack patterns. Our proposed technology, called* Theorem-Carrying-Transaction (TCT)*, combines the benefits of runtime checking and symbolic proof. Under the TCT protocol, every transaction must carry a theorem that proves its adherence to the safety properties in the invoked contracts, and the blockchain checks the proof before executing the transaction. The unique design of TCT ensures that the theorems are provable and checkable in an efficient manner. We believe that TCT holds a great promise for enabling provably secure smart contracts in the future. As such, we call for collaboration toward this vision.*

## 1. Introduction

We stand at the threshold of a decentralization revolution that holds the power to transform our society. However, there are significant obstacles that must be overcome first. One such obstacle is that executing published smart contracts on all blockchain nodes through consensus does not offer enough transparency and trustworthiness. This is because the behavior of contract code is often too complex to be restrained. As a result, there have been numerous high-profile incidents that have resulted in significant financial losses due to unintentional vulnerabilities and intentional logic trapdoors in smart contracts. Unfortunately, because of the tenet "code is law" in the decentralization paradigm, transactions are immutable, and losses are irreversible.

To enhance the security of smart contracts, the conventional approach is to rely on tools and/or human experts to detect as many bugs or known vulnerabilities as possible and resolve them to prevent unintended contract behaviors. In this paper, we propose a distinct approach that enables enforcing *design-level safety properties* that explicitly express intended behaviors of contracts, which are enforced at runtime. To achieve this, we introduce a technology called *Theorem-Carrying-Transaction (TCT)*. With TCT, every transaction must carry a theorem that proves its adherence to the specified safety properties in the invoked contracts. This ensures that transactions that violate any safety property are blocked, regardless of *any* hidden bugs or unexpected vulnerabilities in the deployed code. TCT empowers smart contract code to explicitly declare design-level safety properties, which are guaranteed to hold true by the blockchain VM (EVM or a similar VM). The uniqueness of TCT lies in its design. It combines the benefits of runtime checking and symbolic proof, ensuring that the theorems are provable and checkable in an efficient manner. Once a theorem is validated and accepted by the blockchain, future transactions can reuse it to affirm their safety with minimal runtime overhead. This working draft aims to document the fundamental concepts and technical thoughts of TCT.

### 1.1. Why are design-level properties the key to the trust of smart contracts?

Conceptually, trustworthiness means that a system behaves according to its design specification. However, when dealing with a real-world security problem, mainstream approaches tend to reduce it into a matter of enumerating specific low-level violation patterns. For example, many security issues are categorized as "input sanitization bug", attributing the root cause to programmers' failure to anticipate a specific form of malicious input. This type of afterthought is paradoxical, as it demands programmers to expect every unexpected possibility. In other words, they need to enumerate non-goals, as opposed to specifying goals. Trust is hard to establish in this manner.

Smart contract security is in the same situation. For example, integer overflow and reentrancy, two separate low-level patterns, are often given as representative security bugs. In fact, neither is a security bug per se. The code below is in the Uniswap implementation, which clearly states that integer overflow is desired. As for "reentrancy", it is really a synonym of the terminology "callback function" – a function in the callee module, before returning, calls into (i.e., reenters) the caller module. This pattern is common and necessary in many systems but is suddenly considered as a bug pattern under a new name "reentrancy".

```
75    uint32 blockTimestamp = uint32(block.timestamp % 2**32);
76    uint32 timeElapsed = blockTimestamp – blockTimestampLast;     // overflow is desired
```
(from https://github.com/Uniswap/v2-core/blob/master/contracts/UniswapV2Pair.sol)

**Design-level properties**. Design-level properties will free programmers from the burden and risk of enumerating these low-level bug patterns. In Section 3, we show a vulnerable contract of an ERC-20 token [3], which can be exploited through integer overflow and reentrancy. Let's imagine that time travels back to the year 2016 when nobody knew about these bug patterns. What could programmers do to secure the contract? It turns out that even without knowing any of these patterns, if programmers could specify and enforce the design-level properties in Table 1, neither attack would succeed.

Table 1: Design-level properties for an ERC-20 token contract

$(\forall$`x:address | 0` $\leq$ `balances[x]` $\leq$ `totalSupply)` $\wedge$
$(\sum_x$`(balances[x]) == totalSupply)`

The properties declare that every account balance should be non-negative and the sum of all balances should equal the total supply. Essentially, they simply state the programmers' design goals. If the blockchain could enforce these properties, the attacks exploiting integer overflow and reentrancy would violate these properties, thus be thwarted.

This time-travel thought experiment has its practical relevance – trustworthiness should be and can be established without knowing specific implementation-level bug patterns; instead, design-level properties should become the focus.

**A broad spectrum of design-level properties.** Trust concerns extend beyond just vulnerable contracts to also include malicious contracts that aim to defraud users. Design-level properties ensure that users are protected by a contract's explicitly expressed guarantees. The early history of Uniswap saw a crucial step taken in the form of formally explicating the properties of contracts [5]. It is reasonable to assume that having these explicit properties bolstered users' confidence, contributing to Uniswap's prominence as the leading decentralized exchange on Ethereum.

We imagine many types of design-level properties that can guard a contract's security against unintentional bugs and assure users about its benign intentions. However, today's blockchain runtime has difficulty checking them concretely. TCT enables symbolic verification of these design-level properties at runtime, which makes them feasible and highly efficient to enforce.

- *Contract invariants* – contract invariants like the ones in Table 1 are ensured before and after every transaction. They contain quantifiers, sum of map, etc., so are difficult to check concretely.
- *Precise mathematical relations* – for example, Uniswap's design fundamentally relies on a linear relation for liquidity providers (i.e., $x1/y1 == x2/y2$) and a constant-product relation for traders (i.e., $x1*y2==x2*y2==k$). These design-level properties cannot be concretely checked, because a small transaction fee (e.g., 0.3%) is charged for each transaction. TCT can define and check properties under hypotheses (e.g., the fee percentage is hypothesized to 0%), it performs symbolic reasoning. Similarly, fixed-point arithmetic has imprecision, making it hard to define precise mathematical relations. TCT's symbolic reasoning is not affected by it.
- *The Modifies clause* – oftentimes, a programmer wants to assert that a function only intends to modify a set of variables, and any other modifications are unexpected. The Modifies clause, introduced in the Dafny language, serves this purpose [6]. A high-profile smart contract vulnerability, resulting in stealing of over $30+ million of ether at the time, was due to a code path in the Parity Multisig contract that allowed an arbitrary attacker to modify the contract owner. The incident was analyzed [8]. The path involved low-level details, including "library contract", "delegatecall" and "default function", so the existence of such a path was surprising. Again, imagine time travels back, and the programmers specified the variables expected to be modified by the "default function", which naturally would not include the contract's owner variable. The attack would be thwarted without understanding those low-level details. Note that the Modifies clause is difficult to check concretely, as it would need to check all variables. As another example, using the Modifies clause, an NFT token contract can provably convince users that it is indeed a *soulbound* token (i.e., a non-transferrable token) [7]. Even if the programmers try to hide a secretive code path in the contract to transfer ownership, any transaction executing the path is guaranteed to be rejected.
- *Other assertions believed to be unconditionally true* – When an assertion is checked concretely, it incurs a runtime cost. The programmers are often "frugal" to add assertions and consider properties obviously true as "unnecessary" assertions. Therefore, most contracts only contain sporadic assertions. With TCT, assertions unconditionally true will incur zero runtime overhead, due to its symbolic-proof nature. This will enable the assertion-to-code ratio to increase substantially, which is highly valuable for the trust and security of smart contracts. For example, during a token-swap transaction in Uniswap, some intermediate accounts need to hold tokens temporarily. When the transaction ends, their balances should be zero. Otherwise, the remaining balances will be stolen by any user. A thorough code audit may find this property true for all examined pairs of tokens, but adding a TCT assertion will settle it with high confidence and zero runtime cost. TCT does not require assertions to be about design-level properties. We hope programmers will develop a "lavish" habit of adding whatever assertions believed to be true, at the design and implementation levels.

## 1.2. TCT enables behavioral subtyping for smart contracts

Behavioral subtyping, as formulated by Liskov and Wing [9], has been an aspiration in the object-oriented programming (OOP) community for decades. The subtype requirement is as follows.

```
Let ϕ(x) be a property provable about objects x of type T. Then ϕ(y)
should be true for objects y of type S where S is a subtype of T.
```

Unlike the conventional notion of OOP subtyping, which underpins today's smart contracts and many other real-world systems, behavioral subtyping ensures that a derived class cannot violate properties defined in a base class. The assurance that inheritance implies conformance would be ideal for trust and security. For example, the ERC-20 standard today only specifies variables and functions, but the function behaviors are described in English or implicit in the function names. With behavioral subtyping, the ERC-20 standard can be formally specified as a base class with invariants and function pre-/post-conditions. Suppose we define a base class named `ERC20NonNegativeBalance`, which contains the properties in Table 1, then people can easily trust that all descendant contracts are secure, simply because they see the inheritance relations.

However, with today's conventional subtyping, inheritance does not imply conformance. A well-known example is the base class `Rectangle` and its derived class `Square` [10]. For the base class, the `Width-Height` independence can be proven, but this proven property does not necessarily hold for the derived class. Intriguingly, the assertion in the code "`Rectangle r = Square(); r.setWidth(10); r.setHeight(5); assert(r.area()==50)`" is false, which shows that `Square` is not a behavioral subtype/subclass of `Rectangle`. Ensuring behavioral subtyping is known to be difficult by static program verification techniques but is natural for TCT.

### 1.3. Our vision

TCT requires certain enhancements to the existing Ethereum platform. We plan to prototype an Ethereum-compatible blockchain called TCT-Eth. It can execute all existing smart contracts, and importantly, new contracts that contain TCT-style assertions (including invariants). There are two goals to approach: (1) to attract enough attention from the community to operate TCT-Eth successfully; (2) to demonstrate practicality and high confidence from programmers and users about the TCT technique, so that it can be adopted in real-world deployments. Being Ethereum-compatible also means that the decentralized trust assumption remains unchanged, whether it is proof-of-work, proof-of-stake or proof-of-authority.

We envision a number of changes for the future of smart contracts.

- *"Spec is law of code"*. The status quo of blockchain is code-is-law. However, heavy prices have been paid because of the complexity and lack of transparency of the code. It is necessary that we enable formal specs to confine the code's behaviors. If a contract's spec is ultimately effective, a user agreeing on the contract will mean that he/she agrees on the spec only. This is of course idealistic, but a direction that TCT approaches.
- *Programmers' focus*. Because adding unconditionally true assertions incurs zero runtime cost, programmers will substantially increase the assertion-to-code ratio by focusing on specifying assertions. Behavioral subtyping will become a collaborative effort which programmers at different abstraction levels can contribute to.
- *Clearer business interface for code auditing*. Code auditing is an important service that many companies provide for smart contract developers. The service often combines automatic static code analysis with manual code inspection. TCT offers a new business model with a clearer interface between the auditing company and the client (i.e., the contract developer) – the latter provides the code and the initial spec, and the former defines the final version of the spec and delivers proven theorems that all expected transactions conform to the spec. We believe that this interface will help the client earn higher confidence from end users.

The rest of the paper is organized as follows. Section 2 gives background knowledge about two technologies that inspire TCT. Section 3 uses a vulnerable token contract to explain how a theorem proving design-level properties can thwart real-world attacks. The TCT protocol is described in Section 4. In Section 5, we aim at a more compelling showcase – Uniswap. Section 6 calls for a community collaboration on this vision.

## 2. Background

The core idea of TCT is inspired by two technologies about formal verification of real-world systems – proof-carrying code and concolic testing.

**Proof-carrying code (PCC)**. PCC is designed to allow an OS kernel to safely accept a code module from an untrusted party [11]. The novelty is that the producer of the untrusted module bears the responsibility to prove its conformance to the safety properties declared by the OS kernel. Inspired by this, theorem-carrying transaction (TCT) is a mechanism that requires every transaction to carry a theorem to prove its conformance to safety properties declared by all smart contracts that it invokes. It is worth noting that, unlike PCC, the granularity of a TCT theorem is per transaction, rather than per code module.

**Concolic testing**. The core idea of concolic testing is to perform symbolic reasoning on every concrete execution trace of a complex program. For example, SAGE [13] is a fuzzing technology that has shown successes in exposing a great number of bugs in Microsoft Windows and Office code. SAGE is categorized as a white-box fuzzing technique, because for every input it generates, the program is concretely executed, and the execution trace is recorded. The conjunction of all the branch conditions along the trace forms the "path condition" of the trace. To generate the next test input, the fuzzer negates some of the branch conditions, which forms a new path condition to be presented to a constraint solver. This guarantees that every test input will result in a new code trace, much more effective than a random fuzzing approach. The success of SAGE gives us confidence that a concolic approach is suitable for smart contract executions since the code complexity of a smart contract is much lower than that of Windows and Office x86 executables.

## 3. Example – a token contract with security vulnerabilities

Our example is a contract containing two vulnerabilities. The first is an integer overflow, similar to the one described in ref [1]. The second is a reentrancy vulnerability described in [4]. The contract code is given in Appendix A, and the attack code is given in Appendix B. Figure 1 illustrates the attacks.

### 3.1. Vulnerabilities and attacks

In attack 1, the attacker issues a transaction that calls `transferProxy`. Unexpectedly, the arguments `_value` and `_fee` are set to $2^{255}+1$ and $2^{255}$, respectively. In the line marked as "integer overflow", these two arguments are added, which yields result `1` due to overflow, so the `require` statement does not revert. The rest of the function also follows through. The consequence is that the balances of accounts `msg.sender` and `_to` become enormous, causing an extreme inflation of this token ecosystem.

Attack 2 demonstrates the consequence that the attacker can make the balance of one of his accounts `50` by clearing another account with the balance `5`. It is achieved by making reentrant calls for 10 times as shown in Figure 1. Essentially, when the token contract calls function `receiveNotification` in the attacker's contract, the latter calls function `clear` again in the former. This unexpected code sequence results in "`balances[_to] += bal`" executed 10 times. The first instance of reentrancy attack, dated back to July 2016, was a landmark event in the early history of Ethereum. The attacker stole 3.6 million Ethers. Moreover, it seriously damaged the credibility of the Ethereum platform. A long-term impact of this incident is that the original "Ethereum" community has since been forked into "Ethereum (ETH)" and "Ethereum Classic (ETC)" [15], which was a heavy price to pay.

As explained in the introduction, integer overflow and reentrancy are just two low-level details of the runtime, which are not always bugs or vulnerabilities. The reason why this token contract is vulnerable is because there are unexpected code paths that lead to the violations of the properties in Table 1. These code paths happen to utilize integer overflow and reentrancy.

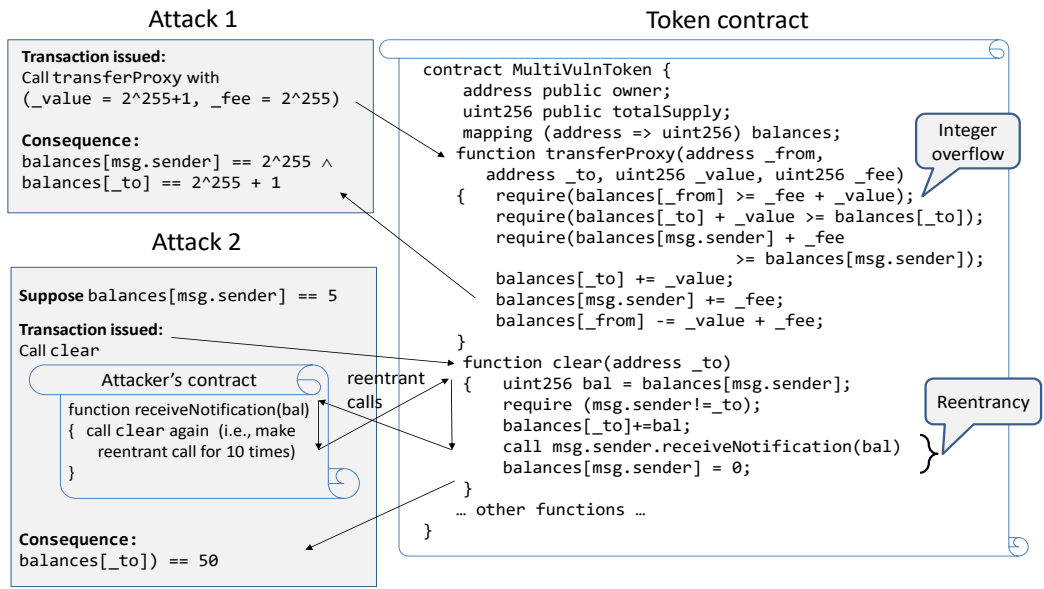

Figure 1: integer overflow and reentrancy attacks exploiting a vulnerable contract.

### 3.2. Proving conformance of a code path as a theorem

We now use formal reasoning to examine why the attacks violate the properties in Table 1. Let's study Attack 1 first. The code path of this transaction is the following straight-line code, which we express using the Boogie language [16].

```
tmp1 := add(_fee , _value);
assume (balances[_from] >= tmp1);

tmp2 := add(balances[_to] , _value);
assume(tmp2 >= balances[_to]);

tmp3 := add(balances[msg.sender] , _fee);
assume(tmp3 >= balances[msg.sender]);

balances[_to] := add(balances[_to] , _value);
balances[msg.sender] := add(balances[msg.sender] , _fee);
balances[_from] := sub(balances[_from] , tmp1);
```

Figure 2: straight-line code path corresponding to Attack 1

Note that there are axioms to specify the basic properties of mathematical operations, including the following. These are not specific to the code we study, but the general truth of math.

```
function sum(m: [address] int256) returns (int256);    //Sum of map
axiom (forall m: [address] int256, a:address, v:int256 :: sum(m[a:=v]) == sum(m)- m[a] + v);
axiom (forall m: [address] int256 :: ((forall a:address :: 0<=m[a])
                                        ==> (forall a:address :: m[a]<=sum(m))));
function add(a,b:int256) returns (int256);      // Modulo-2^256 add
axiom (forall a,b: int256 :: add(a,b) == (a+b) % TwoE256);
function sub(a,b:int256) returns (int256);      // Modulo-2^256 subtract
axiom (forall a,b: int256 :: sub(a,b) == (a-b) % TwoE256);
```

Figure 3: some axioms for basic math operations

The properties in Table 1 are invariants that must hold before and after every transaction. Therefore, the proof obligation can be expressed as follows. Two "inv" assume-statements are added in the beginning of the code path, and two "inv" assert-statements are added in the end.

```
(hyp)   assume (φ(v,s));

(inv)   assume (sum(balances) == totalSupply);
(inv)   assume (forall x:address :: 0 <= balances[x] && balances[x] <= totalSupply);

        tmp1 := add(_fee , _value);
        assume (balances[_from] >= tmp1);
        tmp2 := add(balances[_to] , _value);
        assume(tmp2 >= balances[_to]);
        tmp3 := add(balances[msg.sender] , _fee);
        assume(tmp3 >= balances[msg.sender]);

        balances[_to] := add(balances[_to] , _value);
        balances[msg.sender] := add(balances[msg.sender] , _fee);
        balances[_from] := sub(balances[_from] , tmp1);

(inv)   assert (sum(balances) == totalSupply);
(inv)   assert (forall x:address :: 0 <= balances[x] && balances[x] <= totalSupply);
```

Figure 4: proof obligation of the straight-line code path

The "hyp" line specifies the condition φ(v,s) under which the invariant will hold in the end, where v represents input parameters of the transaction and s represents the blockchain state.

**A theorem**. Suppose the contract programmer realizes that his/her intention of the code is to process a transaction under the condition $0 \leq totalSupply < 2^{255} \land 0 \leq \_value < 2^{255} \land 0 \leq \_fee < 2^{255}$. Thus, the "hyp" line is specified as "`assume (0<=_value && _value<TwoE255 && 0<=_fee && _fee<TwoE255 && totalSupply<TwoE255)`". This Boogie code can be successfully verified, meaning that the following theorem, denoted as τ, has been proven.

Theorem 1:
τ := (MultiVulnToken :: transferProxy, $0 \leq totalSupply < 2^{255} \land 0 \leq \_value < 2^{255} \land 0 \leq \_fee < 2^{255}, 0xa1b2c3d4$)

Assuming that the code path in Figure 2 is uniquely identified by a secure hash value 0xa1b2c4d4, this theorem states that, for a transaction with `MultiVulnToken::transferProxy` as the entry function, if condition φ(v,s) is satisfied and the code path's hash is 0xa1b2c4d4, then all safety properties along the path hold.

It is worth emphasizing that the theorem says nothing about condition ¬ φ(v,s). In Attack 1, $\_value == 2^{255} + 1$, not satisfying φ(v,s). The theorem does not suggest that Attack 1 will violate any safety property. Fundamentally, TCT is not a bug finding technology, but a safety-property enforcement technology. Section 4 will explain the TCT protocol. It requires every transaction to carry a theorem. Because Attack 1 creates balance amounts out of thin air, no theorem can prove the invariants in Table 1, and the transaction will be rejected by the TCT protocol due to the lack of a valid theorem.

Generality of the theorem is also worth emphasizing. Like concolic testing, the code path is concrete (i.e., the one in Figure 2, identified by hash 0xa1b2c3d4) but the variables are symbolic. Section 4, we will explain that the TCT protocol utilizes generalization so that proven theorems are cached and reused, which is important for minimizing the cost.

**Attack 2**. A similar exercise can be done regarding Attack 2. Figure 5 represents the straight-line code. Normally, the two lines marked with "(*)" should be executed once. However, because the attacker utilizes the reentrancy pattern, the computation is repeated 10 times. The programmer may not be aware of the reentrancy pattern, like before the year 2016, but it is the attacker's obligation to prove a theorem why this code path would satisfy the invariants. Of course, the attempt is doomed to fail.

```
    bal := balances[msg.sender];
    assume(msg.sender !=_to));
    tmp1 := add(balances[_to], bal);
    balances[_to] := tmp1;

    /* These are due to reentrancy */
(*) tmp1 := add(balances[_to], bal);
(*) balances[_to] := tmp1;
    …
    tmp1 := add(balances[_to], bal);
    balances[_to] := tmp1;
    /* End of reentrancy */

    balances[msg.sender] := 0;
```

Figure 5: straight-line code path corresponding to Attack 2

## 4. The TCT protocol workflows

In this section, we explain the workflows of the TCT mechanism. Like today's Ethereum system, there are three parties – the transaction issuer (e.g., a browser wallet), a pre-execution service (e.g., *infura.io*) and the Ethereum blockchain. Note that the Ethereum blockchain consists of many nodes. The pre-execution service includes one of the nodes.

The TCT mechanism adds a theorem repository to every Ethereum node, including the node of the pre-execution service. A theorem is in the form $\tau := (f, \varphi(v, s), h)$, where $f$ represents a contract function; φ represents a Boolean condition over input parameters and the blockchain state; and $h$ is the *path hash* as explained earlier. The theorem means that, for any transaction started by invoking $f$ when φ is satisfied, if it is completed (i.e., not reverted by EVM) and the hash of the code path equals $h$, then all the assertions along the execution are guaranteed to hold. Conceptually, the path hash is equivalent to the *path condition* mentioned in Section 2, as both uniquely identify a concrete trace. However, the path hash is simpler to represent and check for a smart contract execution.

**Workflow A.** Figure 6 illustrates a workflow when the transaction issuer tries to submit a transaction Tx to the blockchain. As today, Tx is first submitted to the pre-execution service. The service tries to find a theorem applicable to Tx. If no such theorem is found, the pre-execution service notifies the transaction issuer that it needs to construct and prove a theorem that is expected to be considered applicable to Tx by the blockchain. It is up to the transaction issuer how to achieve this goal. Suppose it defines the theorem τ and constructs its proof ρ. The tuple (Tx, τ, ρ) is submitted to the blockchain. Every node verifies that ρ is a proof of τ and save τ in its theorem repo. Then, the node checks to ensure that τ is considered appliable to Tx w.r.t. the node's state. If so, Tx is executed.

To check ρ, the node first needs to get the EVM code trace by test-running the transaction Tx. The recorded trace is compared to the hash h to confirm the match. With the code trace available, the node checks whether ρ gives a sound reasoning to prove that "under condition φ (v,s), the code trace does not violate any assertion".

Note that, in this workflow, proving τ (i.e., generating ρ) by the transaction issuer and checking ρ by the blockchain node are different mechanisms. This is like the proof-carrying-code (PCC) concept described in Section 2. An alternative approach is to use the same prover on both parties. For example, Boogie can be used as the prover. The transaction issuer proves theorem τ and send it to the blockchain without ρ. The blockchain also uses Boogie to re-prove and accept τ.

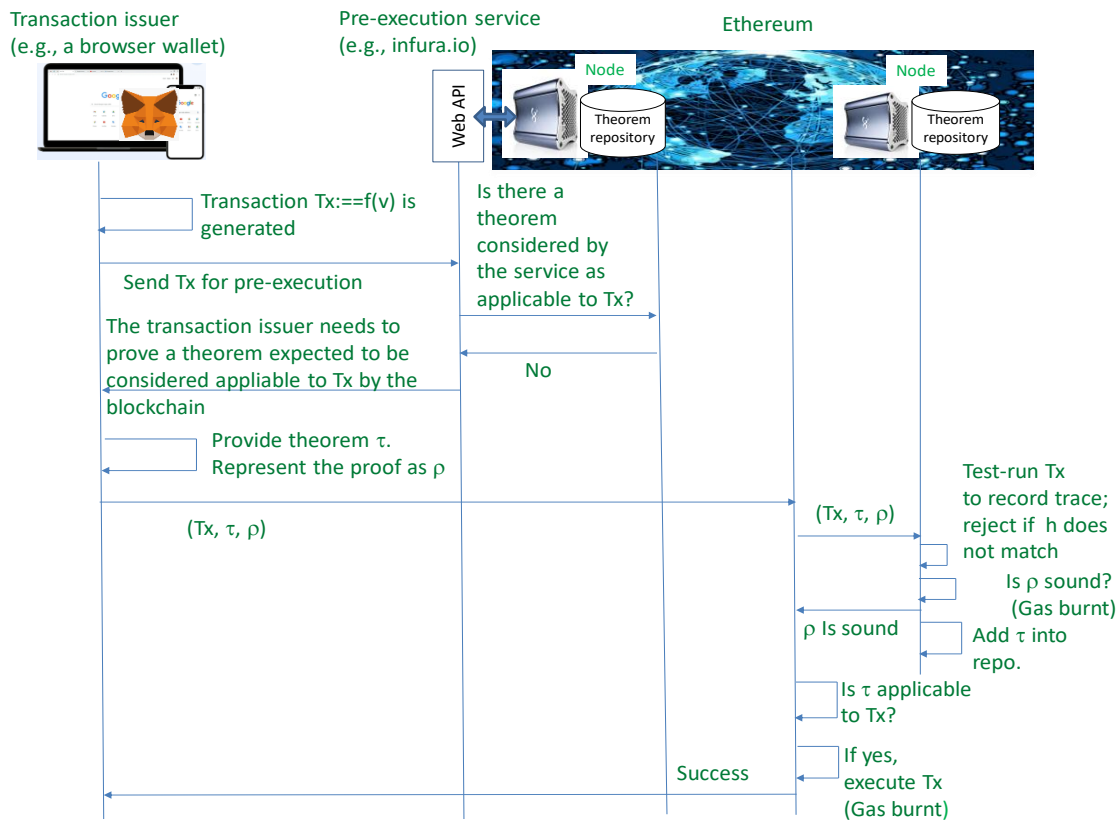

Figure 6: workflow when a transaction has no applicable theorem to prove its security.

**Workflow B.** Figure 7 illustrates a different workflow. This time, the pre-execution service finds a theorem τ that it considers applicable to Tx. The tuple (Tx, τ) is submitted to the blockchain. Every node simply checks the existence of τ in the repo. If so, it follows the rest of the steps described in the previous workflow.

The generality of the theorem is important. Suppose theorem 1 in Section 3.2 is previously proven and accepted for the transaction `MultiVulnToken::transferProxy(_from=…, _to=…, _value=10, _fee=1)`, and the current transaction is `MultiVulnToken::transferProxy(_from=…, _to=…, _value=20, _fee=2)`. In this case, TCT will apply workflow B to reuse the theorem. The only runtime overhead is to look up the theorem in the repo and check the condition "$0 \leq totalSupply < 2^{255} \wedge 0 \leq 20 < 2^{255} \wedge 0 \leq 2 < 2^{255}$" and "$h == 0xa1b2c3d4$".

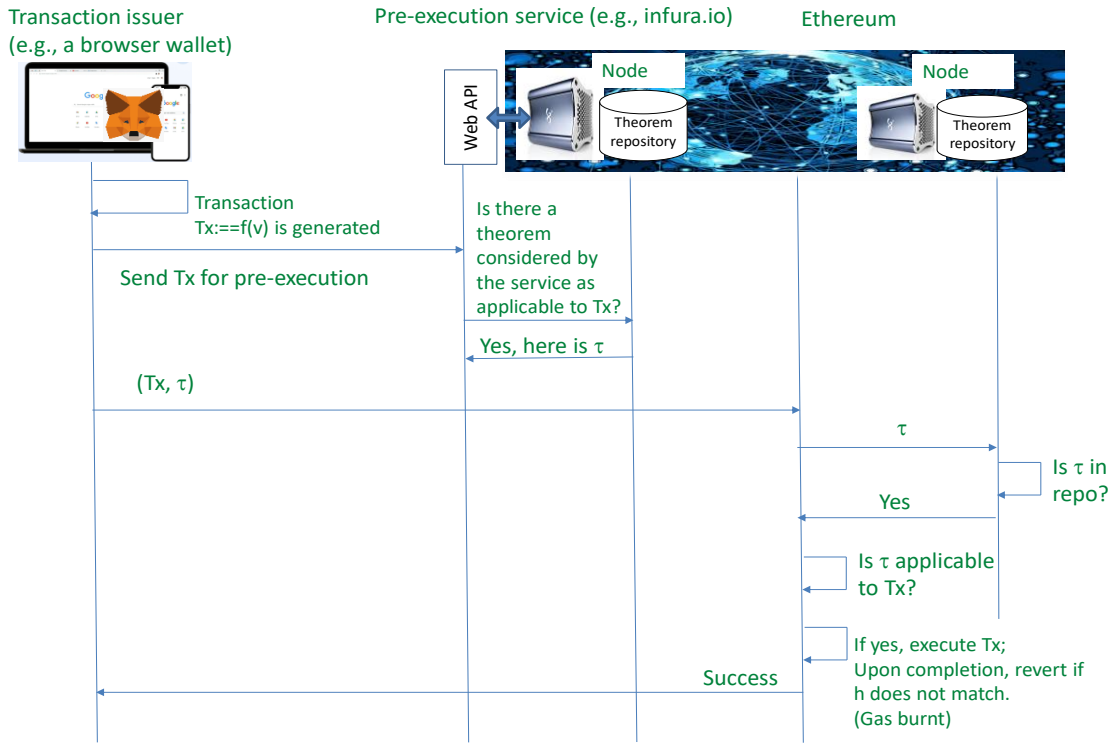

Figure 7: workflow when a transaction has an applicable theorem to prove its security.

**Workflow C**. Besides the above two workflows, any user, typically a programmer of the contract during the testing phase, can submit any theorem τ with its proof ρ to the blockchain as shown in Figure 8. If ρ is sound, τ will be saved into the repo. The only effect is to populate the theorem into the theorem repo. The transaction is not actually committed to the blockchain state.

We expect that most theorems for a contract are accepted by the blockchain via workflow C, because the contract programmers are supposed to run test cases that cover all normal scenarios they can imagine. After the contract starts to process real users' transactions, workflow B should be the primary workflow, because most transactions are expected to be normal. Workflow A happens only when a real user issues a transaction that goes through a code path unseen in all test cases. It may be an attack. If it is benign, the user unfortunately needs to take the burden to provide a new theorem.

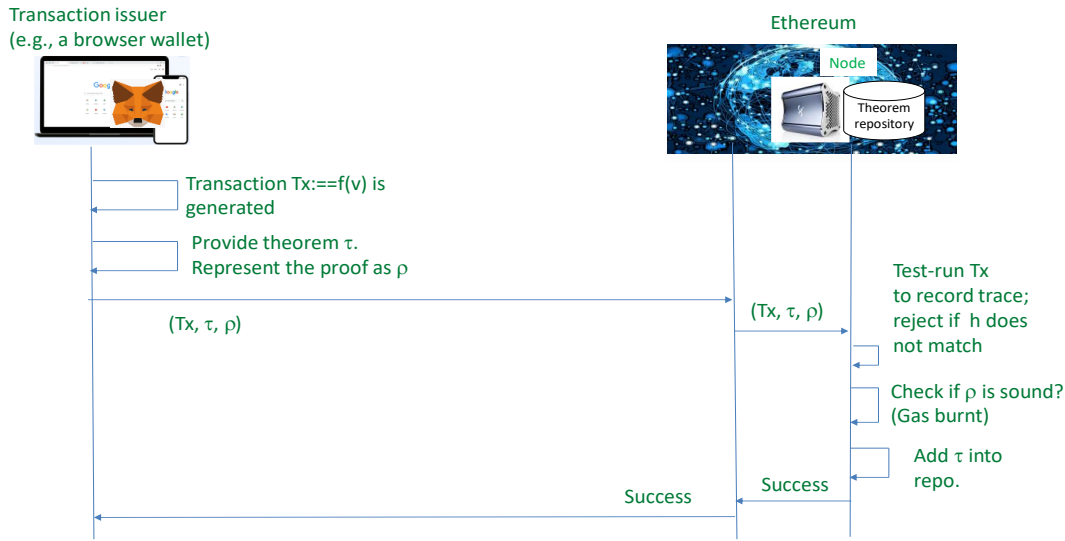

Figure 8: workflow for a programmer to submit a theorem and its proof.

## 5. A more compelling showcase – Uniswap

Section 3 and Section 4 have shown the practicality and value of using TCT for token contracts, which is exciting, as tokens are a major application for Ethereum. To further showcase TCT, we plan to apply it on the Uniswap decentralized exchange system, which is currently one of the most extensively used systems on Ethereum, based on the amount of gas burnt [18]. The complexity of a Uniswap transaction is significantly greater than an individual token transaction, as it often involves two or three tokens.

It is crucial to strengthen the public's confidence in decentralized exchanges since recent events have exposed how the lack of transparency in centralized exchanges has enabled them to manipulate their clients' funds at will, such as in the FTX scandal [17]. As Uniswap operates as a decentralized system, its smart contracts are publicly available. If design-level safety properties can be built into these contracts, more users will feel confident in using the system.

### 5.1. Brief explanation of the protocol

Uniswap implements an automatic market maker (AMM) protocol to allow users holding different ERC20 tokens to form a "market", which dynamically and automatically set prices for these tokens. Assuming most users want to maximize their gains from the trades, the prices will reach equilibriums that represent the fair market prices.

Uniswap has evolved to version V3. We study V2 because it already contains most of the essentials related to the main properties of exchanges. Uniswap V2 protocol is described in refernce [20]. There are two types of users. The first type is Liquidity Provider (LP). LPs create pools as reserves to facilitate trading of different ERC-20 tokens. Every pool is defined by a token pair. In our example, we assume there is a pool of the AAA token and BBB token. The pool's position can be represented in a coordinate system as shown in Figure 9. For example, $(x1, y1)$ means the pool currently holding x1 AAA tokens and y1 BBB tokens. Other than creating pools, an LP can also add or remove AAA tokens and BBB tokens to or from a created pool. The protocol requires that the ratio between the two tokens not changed when an LP performs these operations. In Figure 9, this is shown as the dashed line denoted as "x1/y1=x2/y2", which goes through (0,0) and (x1,y1). The new coordinate that an LP can reach, denoted as (x2,y2), must be on this line (in the first quadrant only, excluding (0,0)). Each pool has its own "liquidity token". At any time, an LP holds a number of liquidity tokens proportional to the percentage of her contribution of AAA tokens relative to the total AAA token reserve in the pool (which equals the percentage of her contribution of BBB tokens relative to the total BBB token reserve in the pool).

The other user type is trader. The pool in Figure 9 facilitates a trader to swap AAA tokens for BBB tokens, or vice versa. Suppose the current coordinate is (x1,y1). The protocol requires the trader to move the coordinate along the hyperbola x1*y1=k1, in which k1 is a constant. For example, (x1',y1') is a valid destination from (x1,y1). Moving to this destination means that the trader gives the pool ∆x1=(x1'-x1) AAA tokens, and receives from the pool ∆y1= (y1-y1') BBB tokens.[1]

When the market is in action, LPs' operations and tranders' operations interleave, so the joint effect is to move the coordinate inside the quadrant, e.g, along the path (x1,y1) -> (x2,y2) -> (x2',y2').

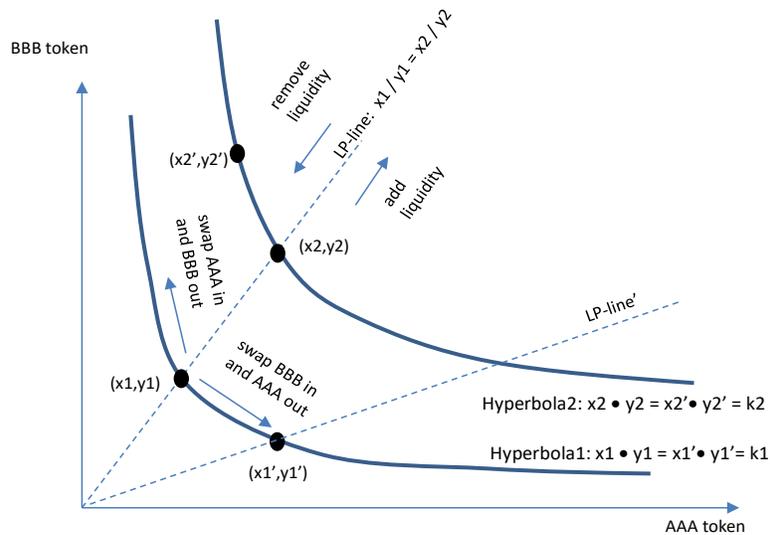

Figure 9: coordinate (x,y) moves upon liquidity and swap operations.

The description above does not take fees into consideration. To incentivize LPs, every swap operation is charged a fee paid as 0.3% of the amount of swap-in tokens. For example, when a trader wants to move the coordinate from (x1,y1) to (x1', y1'), because 0.3% of ∆x1 is taken as the fee, ∆y1 will be slightly smaller than it would in the above no-fee scenario. The detail, which is unnecessary for our further description, is given in Section 1 of reference [19]. Because the fee is added to the pool's reserve, the hyperbola moves slightly upward, because k1 is slightly increased.

## 5.2. Security properties for Uniswap V2

We want the Uniswap V2 contracts to ensure the LPs that their investments are "safe" and ensure the traders that their trades are "fairly processed". The Uniswap scenario is about investment and trade, which involves the market. Unlike for an ERC-20 contract, we cannot demand that "safety" and "fairness" mean no financial loss. What are the meaningful things we can say about the desired properties of the contracts? Intuitively, we can describe some properties as follows.

---

[1] The economic rationale of the protocol is beyond the scope of our description. Basically, if every trader tries to profit from trading, the coordinate will eventually settle in an equilibrium which represents the market's consensus on the relative price between AAA and BBB.

- Every `addLiquidity` operation "works as expected". The meaning is that, assuming no fee, (1) the coordinate move is along the LP line; (2) if an LP is rewarded ∆l liquidity tokens when she calls `addLiquidity` with ∆x AAA tokens and ∆y BBB tokens, she will get ∆x AAA tokens and ∆y BBB tokens back at anytime by calling removeLiquidity with ∆l liquidity tokens, as long as there is no swap-transaction. In other words, many LPs can freely move the coordinate along an LP line in Figure 9. The contracts guarantee that everyone will get all her investments back at anytime she wants, if no trader moves the coordinate along the hyperbola. (Perhaps another way to state the guarantee is that "No LP's action can affect another LP")
- Every swap-transaction "works as expected". The meaning is that, assuming no fee, (1) the coordinate move is along the hyperbola; (2) if a trader trades in ∆x AAA tokens and gets ∆y BBB tokens back, and immediately trades in ∆y BBB tokens, she should get ∆x AAA tokens back.
- Other intuitive properties, such as "the balance of any uninvolved user is not changed by any liquidity or swap transaction", and "when any transaction is done, the balances in all intermediate accounts must be zero (otherwise, these balances would be stolen by any user)", etc.

All these are design-level properties. As we hint in the introduction, many properties are difficult to check concretely. For example, the definitions of these properties assume no fee and make assertions about a hypothetical state after a future transaction happens (e.g., a future `removeLiquidity` will gets back the tokens invested in the current addLiquidity). If they were checked concretely, the hypothetical future transactions would need to be executed for real, which would change the blockchain's concrete state. The symbolic-proof nature of TCT makes it feasible to define and enforce these properties.

## 6. Call for collaboration: building the TCT-Eth blockchain

Our first milestone is to demonstrate a test net of the TCT-Eth blockchain, along with the TCT-capable browser wallet and pre-execution service (shown in Figure 10). They speak the TCT protocol described in Section 4. On this test net, we will demonstrate the end-to-end feasibility of decentralized exchange using the TCT-enhanced Uniswap contracts to trade ERC-20 tokens.

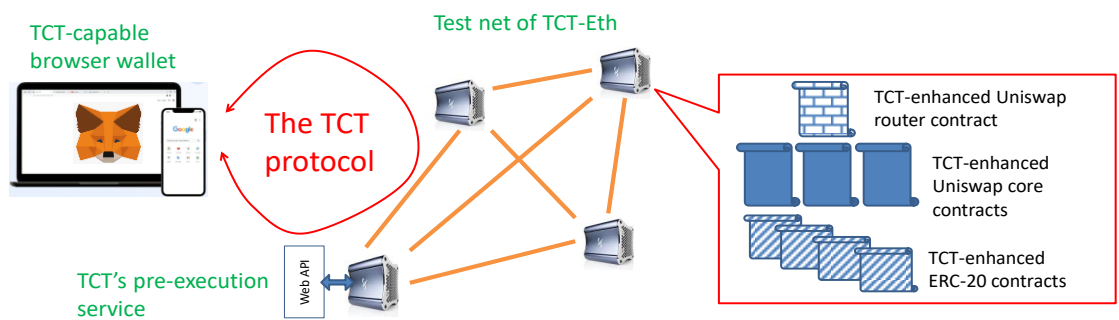

Figure 10: Test net and decentralized exchange as the first showcase of TCT.

### 6.1. Main components and expertise areas

To realize this vision, we need collaborators in the following expertise areas.

**Ethereum**. We need to enhance the Go-Ethereum implementation [26] for TCT. One of our previous projects Forerunner [30] has already built the functionality to record concrete EVM code path and applied optimization techniques on it. It is a nice codebase that the TCT project can base upon. We need to record the code path, reassemble the corresponding Boogie straight-line code, invoke Boogie to verify it, and store it in the theorem repository.

**Solidity and EVM**. We need to establish a mechanism for programmers to specify TCT-style assertions and invariants. Our first goal is to have a design for Solidity code. We can explore using the Solidity metadata feature [23], which allows for comments in the NatSpec format [24] to annotate each contract, interface, function, and so on. Programmers can implement contract invariants and function post-conditions as *TCT-spec functions*, and bind these functions to their corresponding contracts and functions using custom tags such as `@custom:TCT-invariant` and `@custom:TCT-post-condition`.

We also need the expertise of EVM code analysis. The technique may resemble the Halmos EVM symbolic checker [25], but TCT only necessitates the ability to symbolically execute straight-line code.

**Formal verification**. In the first stage, we plan to use Boogie as a verification oracle. The verification focus in this stage is to study the smart contracts in Figure 10. In the second stage, we will explore deeper verification techniques, including precondition generation for linear/non-linear programs, and proof-object generation (i.e., generation of checkable proof). The choice of Boogie for initial prototyping is based on its existence and easiness to integrate text-based intermediary format.

**Wallet and pre-execution service**. We need expertise in browser wallet and the pre-execution service.

## 6.2. Community

A solid partnership is crucial because of the complexity of the project. We need to identify partners (1) who genuinely share the same vision and passion; (2) who have the expertise to contribute; (3) who want to make a generous amount of commitment to this partnership. Fortunately, this situation is common in blockchain/smart contract projects. Examples include the Ethereum foundation [27], the Uniswap community [28] and the Zcash foundation [29], etc.

Incentive is a common mechanism to form a successful partnership. It is an area where we currently lack experience. In addition to a shared vision and technical excitement and desire for scientific advances, we hypothesize the business objectives around security guarantees and the enhanced trust that TCT provides forms a basis for attracting stake holders. We speculate that the emerging industry around formal methods for auditing smart contracts could build business models around pre-certified contracts.

Assuming a strong core team, we will need to incentivize other people to be contributors, including system programmers to work on our open-source code in GitHub, Solidity programmers who write smart contracts, organizations who fund this project, organizations who run blockchain nodes, and early users (similar to early users of Uniswap). A thoughtful design of the incentive contract and associated standardized or supported formats, which will be deployed on the Ethereum main-net, is required.

## References:

[1] An integer overflow vulnerability in the MESH token https://peckshield.medium.com/integer-overflow-i-e-proxyoverflow-bug-found-in-multiple-erc20-smart-contracts-14fecfba2759

[2] Shur number Five. https://arxiv.org/abs/1711.08076

[3] The Ethereum Foundation. ERC-20 Token Standard. https://ethereum.org/en/developers/docs/standards/tokens/erc-20/

[4] Reentrancy Attacks and The DAO Hack. https://blog.chain.link/reentrancy-attacks-and-the-dao-hack/

[5] Uniswap Labs. A short history of Uniswap # Formalized Model. https://blog.uniswap.org/uniswap-history#formalized-model

[6] Jason Koenig and K. Rustan M. Leino. Getting Started with Dafny: A Guide. https://www.microsoft.com/en-us/research/wp-content/uploads/2016/12/krml220.pdf

[7] E. Glen Weyl, Puja Ohlhaver, Vitalik Buterin. Decentralized Society: Finding Web3's Soul. https://papers.ssrn.com/sol3/papers.cfm?abstract_id=4105763

[8] Lorenz Breidenbach, Phil Daian, Ari Juels, and Emin Gün Sirer. An In-Depth Look at the Parity Multisig Bug. https://hackingdistributed.com/2017/07/22/deep-dive-parity-bug/

[9] Barbara Liskov and Jeannette Wing. "A behavioral notion of subtyping". ACM Transactions on Programming Languages and Systems. ACM Transactions on Programming Languages and Systems, Volume 16, Issue 6, 1994. Online version: https://www.cs.cmu.edu/~wing/publications/LiskovWing94.pdf

[10] Stefano Santilli. The Square/Rectangle Enigma. https://www.linkedin.com/pulse/squarerectangle-problem-stefano-santilli/?trk=pulse-article_more_articles_related-content-card

[11] George Necula and Peter Lee. Safe kernel extensions without run-time checking. Proceedings of the USENIX 2nd Symposium on Operating Systems Design and Implementation (OSDI), 1996

[12] Wikipedia. Concolic testing. https://en.wikipedia.org/wiki/Concolic_testing

[13] Patrice Godefroid, Michael Y. Levin, and David Molnar. "Automated Whitebox Fuzz Testing". Network and Distributed System Security (NDSS) Symposium, 2008.

[14] Morgan Peck. Hard-Fork Coming to Restore Ethereum Funds to Investors of Hacked DAO". IEEE Spectrum: Technology, July 19 2016. https://spectrum.ieee.org/hacked-blockchain-fund-the-dao-chooses-a-hard-fork-to-redistribute-funds

[15] The Ethereum Foundation. The history of Ethereum. https://ethereum.org/en/history/

[16] K. Rustan M. Leino. This is Boogie 2. https://www.microsoft.com/en-us/research/wp-content/uploads/2016/12/krml178.pdf

[17] Wikipedia. Bankruptcy of FTX. https://en.wikipedia.org/wiki/Bankruptcy_of_FTX

[18] Ultrasound Money - Burn Leaderboard. https://ultrasound.money/

[19] Yi Zhang, Xiaohong Chen, Daejun Park. Formal Specification of Constant Product (x × y = k) Market Maker Model and Implementation. https://github.com/runtimeverification/verified-smart-contracts/blob/master/uniswap/x-y-k.pdf

[20] Uniswap Docs. Protocol Overview for Uniswap V2. https://docs.uniswap.org/protocol/V2/concepts/protocol-overview/how-uniswap-works

[21] Hayden Adams, Noah Zinsmeister, Dan Robinson. Uniswap v2 Core. https://uniswap.org/whitepaper.pdf

[22] GitHub. Solidity compiler. https://github.com/ethereum/solidity

[23] The Solidity Foundation. Solidity Contract Metadata. https://docs.soliditylang.org/en/v0.8.19/metadata.html

[24] The Solidity Foundation. The NatSpec Format. https://docs.soliditylang.org/en/latest/natspec-format.html

[25] Daejun Park. Halmos. https://github.com/a16z/halmos

[26] GitHub. Go Ethereum. https://github.com/ethereum/go-ethereum

[27] The Ethereum Foundation. "About the Ethereum Foundation." https://ethereum.org/en/foundation/

[28] The Uniswap community. https://uniswap.org/community

[29] The Zcash foundation. https://zfnd.org/about/

[30] Yang Chen, Zhongxin Guo, Runhuai Li, Shuo Chen, Lidong Zhou, Yajin Zhou, Xian Zhang. "Forerunner: Constraint-based Speculative Transaction Execution for Ethereum." ACM Symposium on Operating Systems Principles, October 2021

**Appendix A: An example token contract with integer overflow and reentrancy vulnerabilities**

```
pragma solidity >=0.7.6;

abstract contract Token {
    uint256 public totalSupply;
    function balanceOf(address _owner) public view virtual returns (uint256 balance);
}

abstract contract StandardToken is Token {

    function balanceOf(address _owner) public view override returns (uint256 balance) {
        return balances[_owner];
    }

    mapping (address => uint256) balances;
}

contract MultiVulnToken is StandardToken {
    string public name = "Demo token with reentrancy issue and integer overflow issues";
    UtilLibrary _utilLibrary;
    constructor (uint256 initialSupply) pre post {
        totalSupply = initialSupply;
        balances[msg.sender] = totalSupply;
        require(success);
    }
    function transferProxy(address _from, address _to, uint256 _value, uint256 _fee) public returns (bool){

        require(balances[_from] >= _fee + _value);

        require(balances[_to] + _value >= balances[_to]);
        require(balances[msg.sender] + _fee >= balances[msg.sender]);
        balances[_to] += _value;
        balances[msg.sender] += _fee;

        balances[_from] -= _value + _fee;
        return true;
    }

    //This function moves all tokens of msg.sender to the account of "_to"
    function clear(address _to) public {
        uint256 bal = balances[msg.sender];
        require (msg.sender!=_to);
        balances[_to] += bal;
        bool success;
        (success, ) = msg.sender.call(
            abi.encodeWithSignature("receiveNotification(uint256)", bal)
        );
        require(success, "Failed to notify msg.sender");

        balances[msg.sender] = 0;
    }
}
```

**Appendix B: Exploitations of the vulnerable token contract**

```
//Demo of the attacks
contract reentrancy_attack {
    MultiVulnToken public multiVulnToken;
    address _to;
    uint count = 0;
    constructor (MultiVulnToken _multiVulnToken, address __to)
    {
        multiVulnToken=_multiVulnToken;
        _to = __to;
    }
    function receiveNotification(uint256) public {
        if (count < 9) {
            count++;
            multiVulnToken.clear(_to);
        }
    }
    function attack() public {
        multiVulnToken.clear(_to);
    }
}

contract Demo {
    MultiVulnToken MultiVulnTokenContractAddress;

    address attacker1Address =   address(0x92349Ef835BA7Ea6590C3947Db6A11EeE1a90cFd);
    reentrancy_attack attacker2Address1;
    address attacker2Address2 = address(0x0Ce8dAf9acbA5111C12B38420f848396eD71Cb3E);

    constructor () {
        MultiVulnTokenContractAddress = new MultiVulnToken(1000);
        attacker3Address1 = new
                reentrancy_attack(MultiVulnTokenContractAddress,attacker2Address2);

        //suppose attacker2Address1 has 5 tokens initially
        MultiVulnTokenContractAddress.transferProxy(address(this),
            address(attacker2Address1),5,0);
    }

    function attack1_int_overflow() public {
        MultiVulnTokenContractAddress.transferProxy(address(this),
            attacker1Address,
            uint256(2**255+1),
            uint256(2**255)
        );
    }
    function getBalanceOfAttacker1() view public returns (uint256){
        return MultiVulnTokenContractAddress.balanceOf(attacker1Address);
    }
    function getBalanceOfAttacker2() view public returns (uint256){
        return MultiVulnTokenContractAddress.balanceOf(address(attacker2Address1))
            +  MultiVulnTokenContractAddress.balanceOf(attacker2Address2);
    }
    function attack2_reentrancy() public {
        attacker2Address1.attack();
    }
}
```